\documentclass[pdflatex,sn-mathphys-num]{sn-jnl}% Math and Physical Sciences Numbered Reference Style
\usepackage{graphicx}%
\usepackage{multirow}%
\usepackage{amsmath,amssymb,amsfonts}%
\usepackage{amsthm}%
\usepackage{mathrsfs}%
\usepackage[title]{appendix}%
\usepackage{xcolor}%
\usepackage{textcomp}%
\usepackage{manyfoot}%
\usepackage{booktabs}%
\usepackage{algorithm}%
\usepackage{algorithmicx}%
\usepackage{algpseudocode}%
\usepackage{listings}%
\usepackage{bm}%
\usepackage{float}
\usepackage{natbib}
\usepackage{lineno}
\theoremstyle{thmstyleone}%
\theoremstyle{thmstyletwo}%

\theoremstyle{thmstylethree}%

\begin{document}
%\linenumbers

\title[Article Title]{Projection-shifted particle-flow imaging with cosmic-ray muons}


\author[1]{\fnm{Zibo} \sur{Qin}}%\email{qinzibo@stu.pku.edu.cn}

\author*[1]{\fnm{Qite} \sur{Li}} \email{liqt@pku.edu.cn}

\author[1]{\fnm{Rongfeng} \sur{Zhang}}%\email{zrf430581@stu.pku.edu.cn}

\author[2]{\fnm{Pei} \sur{Yu}}%\email{yupei@gdlhz.ac.cn}

\author[1]{\fnm{Cheng-en} \sur{Liu}}%\email{liuchengen@stu.pku.edu.cn}

\author[2]{\fnm{Liangwen} \sur{Chen}}%\email{chenlw@impcas.ac.cn}

\author[3]{\fnm{Feng} \sur{Zhang}}%\email{zhangfeng8813@foxmail.com}

\author*[1]{\fnm{Qiang} \sur{Li}}\email{qliphy0@pku.edu.cn}

\affil[1]{\small \orgdiv{State Key Laboratory of Nuclear Physics and Technology}, \orgname{School of Physics, Peking University}, \orgaddress{\city{Beijing}, \postcode{100871}, \country{China}}}

\affil[2]{\small \orgdiv{Institute of Modern Physics}, \orgname{Chinese Academy of Sciences}, \orgaddress{\city{Lanzhou}, \postcode{730000}, \country{China}}}

\affil[3]{\small \orgdiv{National Key Laboratory of Plasma Physics}, \orgname{Laser Fusion Research Center (LFRC), China Academy of Engineering Physics (CAEP)}, \orgaddress{\city{Mianyang}, \postcode{621900}, \country{China}}}


\abstract{Cosmic-ray muons are natural probes for non-destructive imaging, but reaching sub-millimetre resolution with realistic exposure times has long been hindered by a fundamental limitation: the stochastic nature of multiple Coulomb scattering defies deterministic reconstruction of particle trajectories inside matter. Here we introduce Projection-shifted MUon transMission tomogrAphy (P$\mu$MA). Rather than localizing individual scattering points, P$\mu$MA records how material perturbations statistically shift the projected transmission tracks of a muon flux onto a virtual imaging plane. Near density boundaries, asymmetric projection-shift statistics naturally produce steep undershoot-overshoot profiles, sharpening edges without additional filtering. Crucially, the method operates robustly even with only two detector planes, a configuration where conventional scattering tomography cannot be applied. Cosmic-ray simulations with a lead knife-edge target yield edge widths as narrow as 1.196 mm, while monoenergetic-beam simulations reach 48 $\mu$m. Using a prototype system, we resolve 2-mm copper letters within two days---a feat unattainable by standard approaches under equivalent conditions. The projection-shift concept can be extended to accelerator- or laser-driven muon beams and to other ions, establishing a generalisable strategy for high-resolution particle-flow imaging.
}




\maketitle

\section{Introduction}\label{sec1}

Cosmic-ray muons, with an average energy of about 3-4 GeV and a flux of approximately 1 cm$^{-2}$ min$^{-1}$ at sea level \cite{Zyla2020}, offer a natural, passive probe capable of penetrating dense structures non-destructively. For decades, muon imaging has relied on two primary modalities: muon transmission radiography (MTR), which measures flux attenuation, and muon scattering tomography (MST), which exploits multiple Coulomb scattering to map material density \cite{Borozdin2003,Bonomi2020}. While MTR has illuminated large-scale targets such as pyramids \cite{Alvarez1970,Morishima2017}, volcanoes \cite{Nagamine1995,D'Errico2020} and large-scale infrastructures \cite{Qi2026}, and MST has proven effective for security screening \cite{Bonomi2020,Valencia2025}, both approaches have struggled to achieve millimeter-scale spatial resolution in realistic cosmic-ray conditions. 

In recent years, many MST algorithms have been developed \cite{Wen2023,Yu2024,Wu2026}, but most of them reconstruct scattering locations from independently estimated incoming and outgoing tracks. However, multiple Coulomb scattering is intrinsically stochastic: the particle trajectory inside matter is not directly observable and generally cannot be uniquely reconstructed. The "scattering point" in conventional MST is therefore a mathematical construct, not a physical observable. In the small-angle regime dominant for cosmic-ray muons, this limitation can lead to unstable localization and blurred material boundaries.

Here we present Projection-shifted MUon transMission tomogrAphy (P$\mu$MA), a framework that seamlessly integrates transmission and scattering information in a fundamentally new way. Rather than trying to localize individual scattering points, P$\mu$MA measures how matter statistically perturbs the transmission of particle flow. Straight transmission tracks are constructed, and all material-induced angular deviations generate measurable projection shifts onto a virtual imaging plane positioned near the sample, converting microscopic angular deflections into measurable in-plane displacements. Near density boundaries, asymmetric projection-shift statistics naturally produce enhanced edge contrast and sharp localization. 

Because the method does not require explicit scattering-point reconstruction, it remains applicable even in minimal two-detector configurations where conventional MST fails. Using
simulations and experiments, we show that P$\mu$MA achieves substantially improved edge localization and millimeter-scale imaging performance under realistic cosmic-ray conditions.

Beyond cosmic-ray applications, the projection-shift concept offers broader significance. It is readily extensible to accelerator-produced \cite{Xu2025,LiuF2025,Cook2017} or laser-generated \cite{Zhang2025} muon beams, as well as to other penetrating particles such as protons or ions, opening new avenues in plasma diagnostics, accelerator physics and high-resolution non-destructive testing across disciplines. 

\section{The P$\mu$MA framework}\label{sec2}

\begin{figure}[H]
\centering
\includegraphics[width=0.99\textwidth]{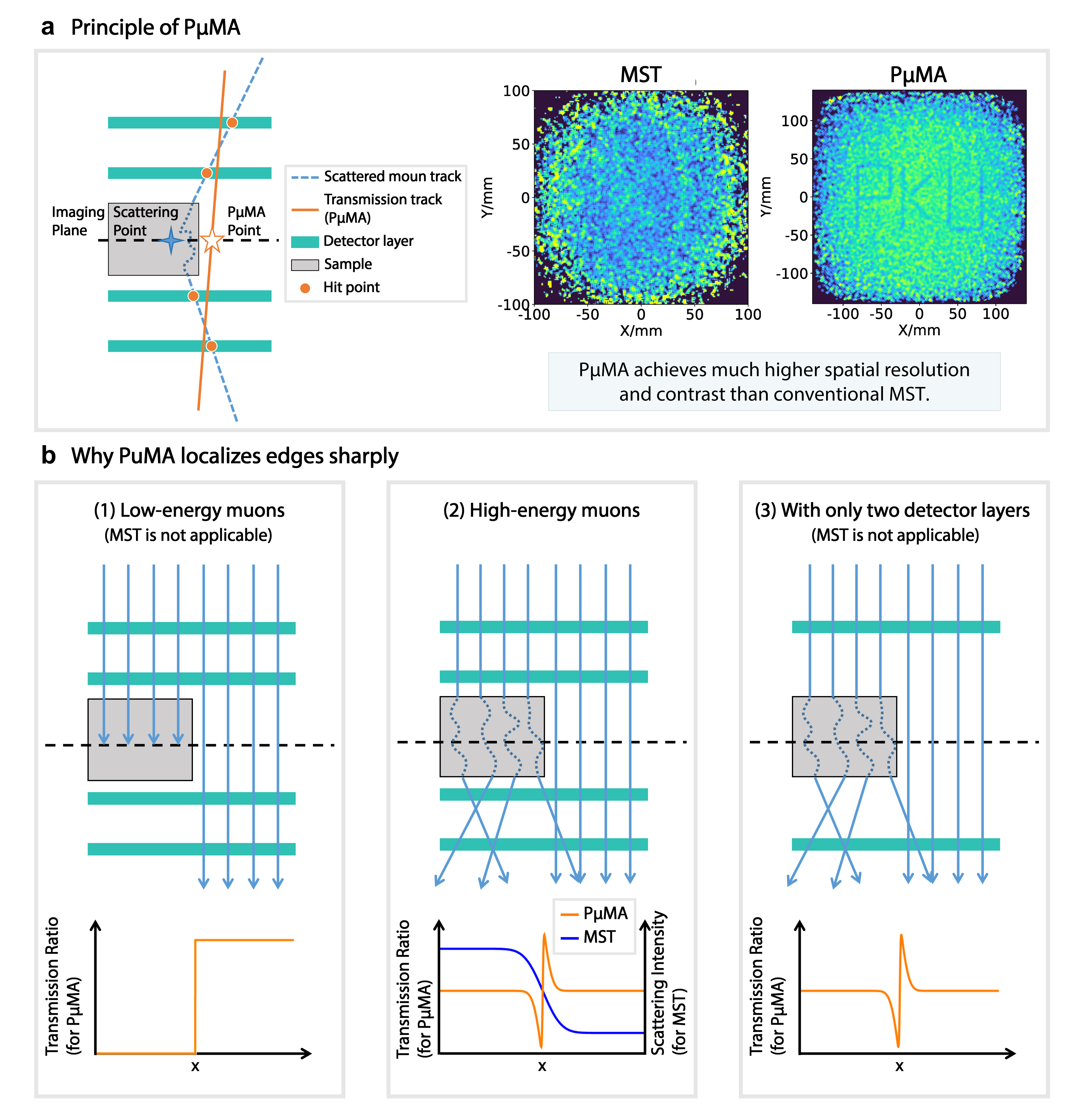}
\caption{\textbf{a}, Principle of P$\mu$MA. Instead of reconstructing scattering points, a straight transmission track is formed by connecting upstream and downstream detector hits. Material-induced angular deflection shifts the intersection of this track with a virtual imaging plane, converting microscopic scattering into a measurable in-plane displacement. The image comparison (right) shows that P$\mu$MA clearly resolves a 2-mm-thick "PKU" copper pattern, whereas conventional MST fails to produce visually discernible features. \textbf{b}, Three scenarios demonstrating P$\mu$MA's robustness and edge sharpness. (1) Low-energy muons stop inside the object: no scattering point can be reconstructed, yet the transmission ratio still yields a sharp boundary. (2) Higher-energy muons: asymmetric projection shifts at material edges produce characteristic undershoot-overshoot profiles, giving enhanced edge sharpening. Methods based on the single-scattering approximation struggle to achieve comparable edge response. (3) Two-detector configuration: scattering points cannot be reconstructed, but P$\mu$MA continues to provide a clear response curve.}\label{fig1}
\end{figure}

Conventional scattering tomography, such as the PoCA algorithm \cite{Bonomi2020}, requires at least four detectors in total (two or more upstream and two or more downstream) to separately measure the incoming and outgoing tracks as independent vectors, from which the scattering angle and an estimated scattering point are derived. In contrast, P$\mu$MA replaces this paradigm with a single geometric construction: a straight transmission track formed by connecting the hit positions in the upstream and downstream detectors. The intersection of this track with a virtual imaging plane placed near the sample defines a Transmission Point (TP) , also referred to as a P$\mu$MA point. When a muon is deflected by matter, the transmission track rotates slightly, shifting the TP laterally in the imaging plane. This lateral displacement, which we call the projection shift, converts the statistical distribution of microscopic scattering angles directly into a measurable in-plane displacement distribution---without ever localizing a scattering point.

Near material boundaries, projection shifts become asymmetric: more TPs are redistributed from high-density regions into neighboring low-density regions than vice versa. This generates a characteristic undershoot-overshoot edge response in the transmission-ratio distribution, producing enhanced edge sharpening and contrast. Fig.~\ref{fig1} contains some schematic diagrams of the P$\mu$MA framework. 

The statistical distribution of the projection shifts, combined with the attenuation of the muon flux (expressed as a transmission ratio), encodes both the areal density and the sharp spatial features of the object. This distribution directly captures how the object alters the muon flow. The P$\mu$MA approach fundamentally changes both the requirements and the paradigm for cosmic-ray muon imaging. 

The P$\mu$MA framework encompasses specific methods, such as P$\mu$MA2, P$\mu$MA4, and P$\mu$MA4C. The number indicates the number of detector planes used in the method, and "C" denotes the application of scattering-angle capping. Detailed descriptions of these methods are provided in Appendix~\ref{secA2}.

\section{Setups}\label{sec3}

\subsection{Experimental setup}\label{subsec1}

This study is a sub-project of the Peking University Muon experiment (PKMu experiment) \cite{YuX2024, Liu2026, Liu2026-2}, which focuses on muon imaging and dark matter searches. The detection system consists of four resistive plate chambers (RPC0-RPC3). (P$\mu$MA2 uses only RPC1 and RPC2.) The RPCs are vertically stacked along the laboratory z-axis, which is defined to be upward and perpendicular to the detector planes. The x and y axes lie in the detector plane and are aligned with the two orthogonal readout-strip directions, with their origin at the center of each RPC’s sensitive area.

The z coordinates of RPC0-RPC3 (top to bottom) are +450 mm, +250 mm, -250 mm, and -449 mm, each known to within 1 mm. Each RPC provides a 280 × 280 $ \mathrm{mm}^{2} $ sensitive area and delivers X/Y positional signals and a timing (T) signal for triggering \cite{Li2012,Li2013,Chen2014}. The X/Y positions are reconstructed using the inductor-capacitor (LC) delay-line readout scheme. The spatial resolution of a single RPC is approximately 0.7 mm, and the two-dimensional reconstruction efficiency exceeds 80\%.

The imaging sample was placed between RPC1 and RPC2 on a 150-mm-high plastic support mounted on RPC2. The z coordinate of the sample’s bottom surface is -73 mm, known to within 1 mm.

\subsection{Simulation setup}\label{subsec2}

Monte Carlo simulations were performed using the Geant4 11.1.2 toolkit \cite{Agostinelli2003,Allison2006,Allison2016}. The detector geometries were provided by the PKMu collaboration and are publicly available on GitHub \cite{PKMUON2024}. For each simulated cosmic-ray event, the horizontal hit position on an RPC was obtained by the energy-deposition-weighted average of all contributing particle coordinates and subsequently smeared with a Gaussian of width $\sigma = 0.7$ mm to match the measured detector resolution. The z positions of the RPCs were fixed to their experimentally determined locations.

Cosmic-ray showers were generated using the CRY package \cite{Hagmann2007}, configured to reproduce the sea-level energy spectrum and angular distribution in Beijing. A planar source of size 300 × 300 $\mathrm{mm}^{2}$---identical to the geometrical footprint of the RPCs---was placed immediately above the uppermost RPC in use. The RPC sensitive area in simulation was kept consistent with the 280 × 280 $\mathrm{mm}^{2}$ area used in the experiment. All relevant electromagnetic and hadronic processes describing the interactions of cosmic rays with the detectors and samples were included. 

In addition to the cosmic-ray simulations, a monoenergetic, parallel muon beam was simulated as an idealized configuration to investigate the ultimate imaging resolution achievable at a representative muon energy. The muon kinetic energy was set to 4 GeV, comparable to the mean energy of cosmic-ray muons at sea level and within the energy range accessible to several muon beam facilities \cite{Xu2025}. The particle generation plane was identical to that used in the cosmic-ray simulations, and the beam direction was perpendicular to all RPC layers. To probe the intrinsic resolution limit at this energy, detector resolution effects were not included in the parallel-beam simulations. 

\section{Results}\label{sec4}

\subsection{Determination of spatial resolution metrics from cosmic-ray muon imaging simulations}\label{subsec3}

To quantitatively compare the spatial resolution of different imaging approaches, simulations of cosmic-ray interactions with a knife-edge sample were performed using the setup in Section~\ref{subsec2}. Standard PoCA-based MST and the optimized MSTC variant \cite{Yu2024} were evaluated using the same simulated dataset as the 4RPC implementation of P$\mu$MA. To suppress statistical fluctuations, large data volumes were generated: 53,802,418 TPs with sample and 57,425,100 without sample for the 2RPC configuration, and 45,966,686 IPs, 37,697,953 TPs, and 20,086,591 post-capping TPs for the 4RPC configuration.

\begin{figure}[H]
\centering
\includegraphics[width=0.9\textwidth]{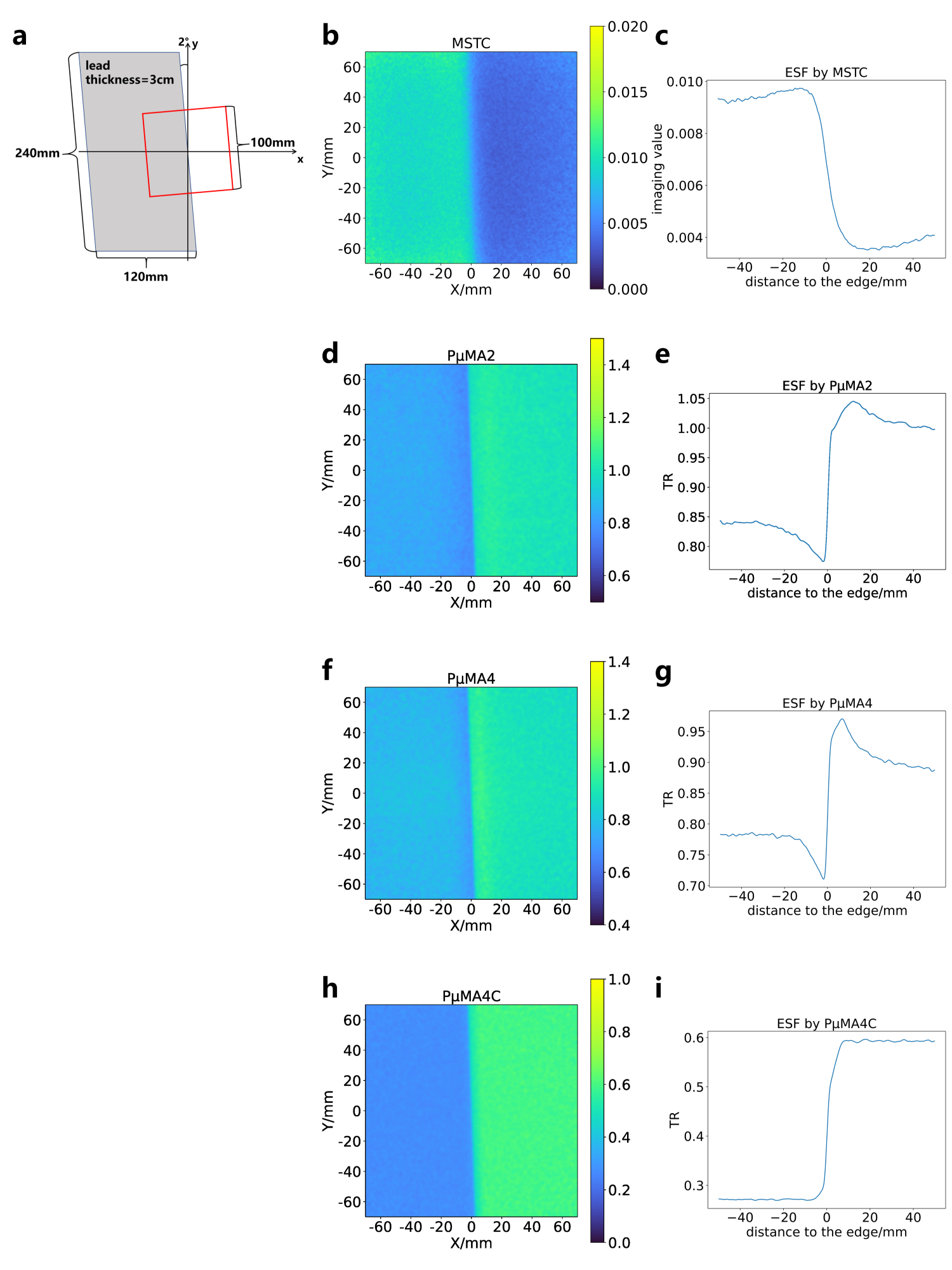}
\caption{\textbf{a}, Schematic diagram of simulated knife-edge sample placement. The area framed by the red line is the sampling area for the edge spread functions (ESFs). \textbf{b}, \textbf{d}, \textbf{f}, \textbf{h}, The imaging results of MSTC, P$\mu$MA2, P$\mu$MA4, and P$\mu$MA4C within the region where -70 mm $<$ x $<$ 70 mm and -70 mm $<$ y $<$ 70 mm. The imaging plane for P$\mu$MA methods is at z = -58 mm. \textbf{c}, \textbf{e}, \textbf{g}, \textbf{i}, The ESFs obtained by MSTC, P$\mu$MA2, P$\mu$MA4, and P$\mu$MA4C.}\label{fig2}
\end{figure}

The sample consists of a right parallelogrammic lead prism. Its base has long and short sides of 240 mm and 120 mm, respectively; the long sides are tilted by $2^{\circ}$ relative to the y-axis, while the short sides align with the x-axis. The origin of the x-y coordinate system is placed at the midpoint of one long side, which defines the knife edge used for analysis (Fig.~\ref{fig2}a). The block thickness is 30 mm, centered at z = -58 mm. 

The parameter settings for each imaging method are as follows: 

\begin{enumerate}
    \item \textbf{MST/MSTC Methods}:
    \begin{itemize}
        \item Smoothing: 3 × 3 pixels; 
        \item Pixel size: 0.5 mm × 0.5 mm; 
        \item Capping angle (MSTC only): $4^{\circ}$; 
    \end{itemize}
    
    \item \textbf{P$\mu$MA Methods}:
    \begin{itemize}
        \item Smoothing (TP without sample or IP): 325 pixels (pixelated circular region, the same below); 
        \item Smoothing (TP with sample): 21 pixels; 
        \item Imaging height: z = -58 mm; 
        \item Pixel size: 0.5 mm × 0.5 mm; 
        \item Capping angle (P$\mu$MA4C only): 0.015 rad. 
    \end{itemize}
\end{enumerate}

For P$\mu$MA methods, strong smoothing of the TP-without-sample or IP matrices is used solely to suppress statistical noise, at spatial scales far larger than the resolution limit, and therefore does not broaden the edge response. The TP-with-sample smoothing is chosen to yield comparable or higher PSNR than MST/MSTC, ensuring a common noise level. As this smoothing inevitably degrades resolution, the resulting comparison is conservative. 

The square region of interest (side length 100 mm), outlined in Fig.~\ref{fig2}a, is selected for quantitative analysis. The reconstructed images by MSTC, P$\mu$MA2, P$\mu$MA4, and P$\mu$MA4C are shown in Figs.~\ref{fig2}b, \ref{fig2}d, \ref{fig2}f, and \ref{fig2}h. Grayscale values within this region are averaged along the knife edge to obtain one-dimensional edge spread functions (ESFs), shown in Figs.~\ref{fig2}c, \ref{fig2}e, \ref{fig2}g, and \ref{fig2}i. The imaging value in MSTC increases with the scattering angle and thus becomes larger within the lead region, opposite to the behavior observed in TR.

Knife-edge widths (20-80\%), spatial resolutions (MTF10), and PSNRs extracted from the ESFs are summarized in Table~\ref{tab1}. Details of these metrics are provided in Appendix~\ref{secA3}. 

In addition, a zenith-angle cut of $2^{\circ}$ is applied to assess the impact of restricting near-vertical muons. For P$\mu$MA2, the cut is applied to transmission tracks; for P$\mu$MA4/4C, it is applied to incident tracks. The corresponding results are listed in Table~\ref{tab1} with a "-Z" suffix. The cut reduces the available statistics to below 5\% of the full dataset, significantly degrading PSNR. Under such noise conditions, reliable MTF10 extraction is not possible for P$\mu$MA2-Z and P$\mu$MA4-Z, whereas knife-edge widths for all methods and the MTF10 for P$\mu$MA4C-Z remain measurable and provide a meaningful basis for comparison.

\begin{table}[h]
\caption{Image quality assessment metrics for each imaging method}\label{tab1}%
\begin{tabular}{@{}llll@{}}
\toprule
Imaging Method & Knife-edge Width (mm) & Spatial Resolution (mm) & PSNR (dB) \\
\midrule
MST         & $7.620\pm0.752$ & $*$           & $34.12\pm0.24$ \\
MSTC        & $7.242\pm0.112$ & $9.11\pm0.30$ & $44.95\pm0.07$ \\
P$\mu$MA2   & $2.128\pm0.108$ & $2.64\pm0.29$ & $42.52\pm0.09$ \\
P$\mu$MA4   & $1.732\pm0.032$ & $2.48\pm0.21$ & $44.72\pm0.07$ \\
P$\mu$MA4C  & $3.769\pm0.048$ & $2.80\pm0.10$ & $50.41\pm0.04$ \\
P$\mu$MA2-Z & $1.315\pm0.138$ & $*$           & $28.17\pm0.48$ \\
P$\mu$MA4-Z & $1.196\pm0.060$ & $*$           & $33.82\pm0.25$ \\
P$\mu$MA4C-Z& $1.308\pm0.064$ & $2.38\pm0.23$ & $34.31\pm0.24$ \\
\botrule
\end{tabular}
$*$ Not measurable due to excessive noise.  
\end{table}

As shown in Table~\ref{tab1}, the P$\mu$MA methods achieve smaller knife-edge widths and higher spatial resolutions than MST/MSTC, demonstrating superior resolving power at the millimeter scale. Imposing a zenith-angle restriction on incident tracks yields further reductions in knife-edge width and additional resolution gains. Although such a constraint is not practical for cosmic-ray-based imaging, the results indicate that improved collimation of incident tracks particularly benefits the P$\mu$MA framework. This observation points to the potential applicability of P$\mu$MA methods in parallel muon-beam imaging settings. 

\subsection{Determination of spatial resolution metrics from monoenergetic parallel muon-beam simulations}\label{subsec4}

To quantify the intrinsic spatial-resolution limit of the P$\mu$MA framework under idealized conditions, simulations were carried out using a monoenergetic, parallel muon beam incident on a knife-edge sample. The simulation setup follows that described in Section~\ref{subsec2}. 

The sample geometry is identical to that used in the cosmic-ray simulations. For the 2RPC configuration, a total of 54,828,790 TPs with sample are obtained. For the 4RPC configuration, 10,800,151 IPs and 10,762,523 TPs are recorded; after applying a scattering-angle cap of 0.005 rad, 5,650,571 TPs remain for image reconstruction. 

To ensure that the extracted knife-edge widths reflect the intrinsic imaging performance rather than discretization effects, very small pixel sizes are employed: 0.01 mm × 0.01 mm for the 2RPC configuration and 0.05 mm × 0.05 mm for the 4RPC configuration. In both cases, the pixel edge length is less than one quarter of the corresponding knife-edge width. No spatial smoothing beyond the pixel scale is applied to the TP matrices. 

Background reference matrices are constructed differently for the two configurations. In the 2RPC case, the background matrix is obtained by averaging the TP counts over an air region defined by 30 mm $\le$ x $<$ 80 mm and -50 mm $\le$ y $<$ 50 mm, with the resulting mean count assigned uniformly to all pixels. In the 4RPC case, the IP counts are averaged directly on a per-pixel basis to form the background reference. In both configurations, this procedure is equivalent to applying a mean smoothing operation with an effectively infinite spatial extent, which is appropriate for a uniform parallel beam. 

\begin{figure}[H]
\centering
\includegraphics[width=0.6\textwidth]{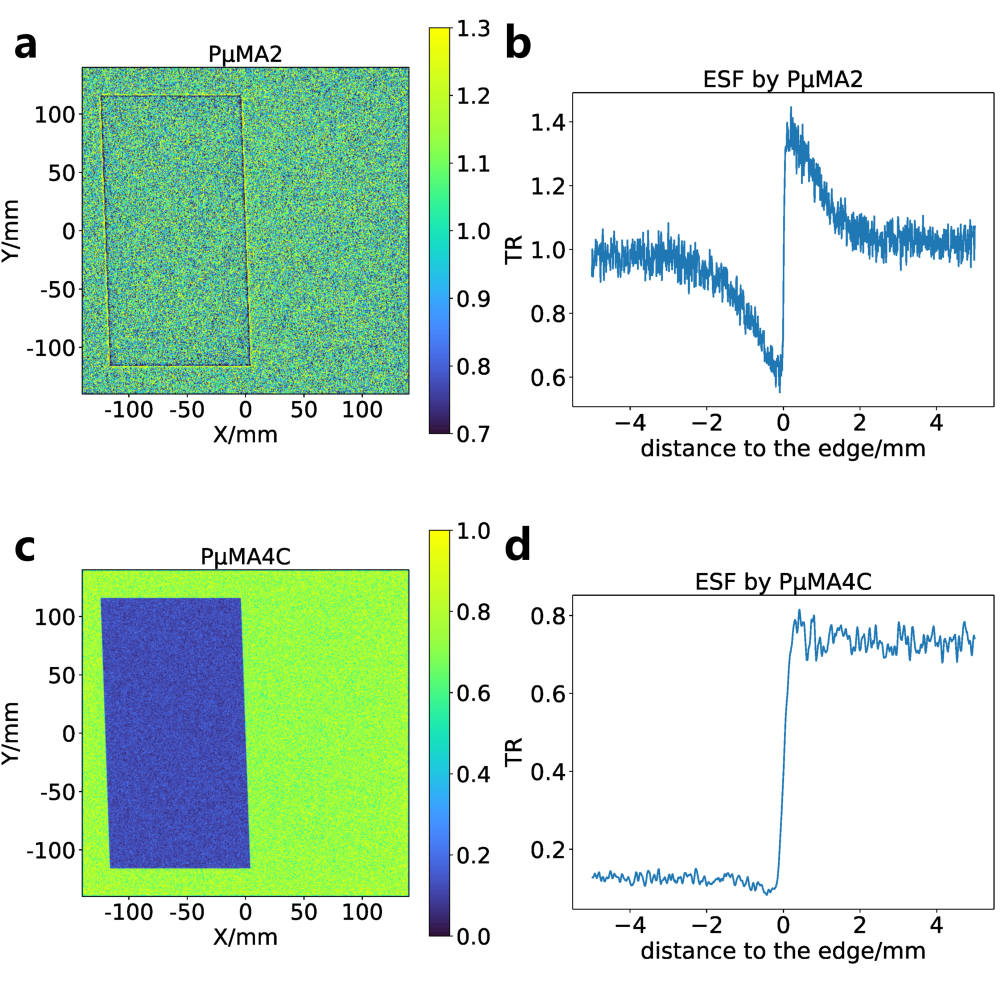}
\caption{\textbf{a}, \textbf{c}, The imaging results of P$\mu$MA2 and P$\mu$MA4C under a monoenergetic, parallel muon beam. The imaging plane is at z = -58 mm. \textbf{b}, \textbf{d}, The ESFs obtained by P$\mu$MA2 and P$\mu$MA4C.}\label{fig3}
\end{figure}

Representative reconstructed images for P$\mu$MA2 and P$\mu$MA4C are shown in Figs.~\ref{fig3}a and \ref{fig3}c, respectively, with the corresponding ESFs presented in Figs.~\ref{fig3}b and \ref{fig3}d. The region of interest used to extract the ESFs is aligned with the knife edge and has the same extent along the edge direction as in the cosmic-ray simulations (100 mm), while its width in the direction normal to the edge is reduced to 10 mm, corresponding to 5 mm on each side of the edge. 

From the ESFs, the knife-edge widths for P$\mu$MA2, P$\mu$MA4, and P$\mu$MA4C are determined to be $0.048\pm0.005$ mm, $0.238\pm0.021$ mm and $0.226\pm0.022$ mm, respectively. Due to the extremely small pixel sizes, achieving ESFs with sufficiently low statistical noise requires very large simulated data sets, which precludes a reliable extraction of the spatial resolutions (MTF10). Nevertheless, based on the cosmic-ray simulation results, the spatial resolution is expected to be of the same order of magnitude as the knife-edge width. 

Overall, the monoenergetic parallel-beam simulations demonstrate the potential for muon imaging to achieve micrometer-scale spatial resolution under beam conditions. In practice, attaining spatial resolutions at the level of tens to hundreds of micrometers requires detector position resolutions of comparable precision. Further improvements may be realized by increasing the muon energy and by minimizing the material budget of the detector system to reduce additional scattering outside the sample region. 

\subsection{Experimental imaging performance for 2D structures}\label{subsec5} 

\begin{figure}[H]
\centering
\includegraphics[width=0.9\textwidth]{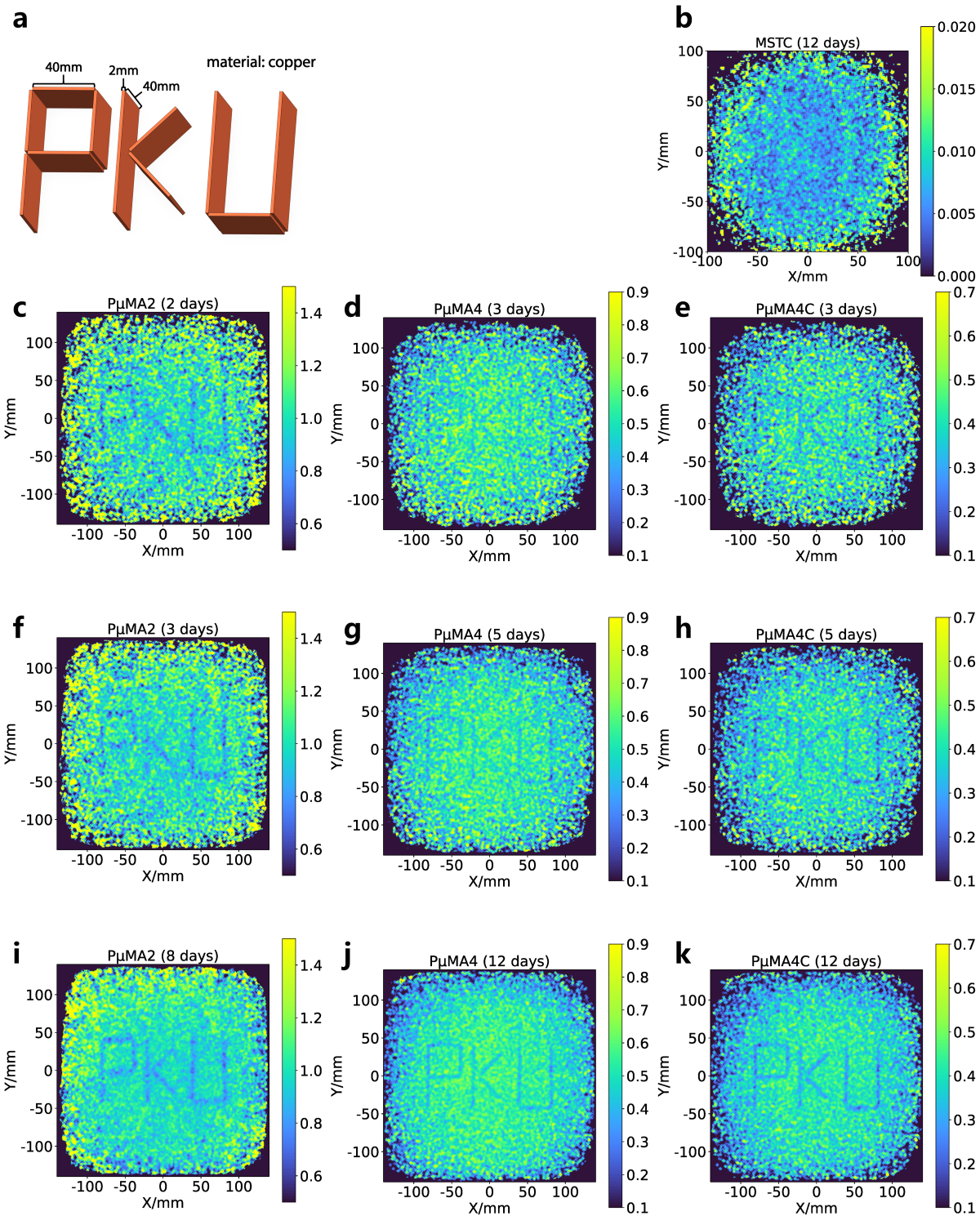}
\caption{\textbf{a}, Schematic diagram of the experimental imaging sample. \textbf{b}, The imaging result of MSTC in about 12 days. \textbf{c}, \textbf{f}, \textbf{i}, The imaging results of P$\mu$MA2 in about 2, 3, and 8 days. (TR is normalized.) \textbf{d}, \textbf{g}, \textbf{j}, The imaging results of P$\mu$MA4 in about 3, 5, and 12 days. \textbf{e}, \textbf{h}, \textbf{k}, The imaging results of P$\mu$MA4C in about 3, 5 and 12 days. }\label{fig4}
\end{figure}

To assess the millimeter-scale imaging performance of the P$\mu$MA algorithms, a set of copper sheets (40 mm × 40 mm × 2 mm) arranged to form the letters "PKU" (Fig.~\ref{fig4}a) was mounted vertically on plastic supports and imaged for 8 days in the P$\mu$MA2 configuration and 12 days in the P$\mu$MA4 configuration using the setup described in Section~\ref{subsec1}. For P$\mu$MA2, an additional background run without the sample was acquired, although this is not strictly required because a sample-free TP matrix can also be generated by simulation. The center height of the copper target is z = -51 mm.

Reconstruction results obtained with P$\mu$MA2 over 2, 3, and 8 days, and with P$\mu$MA4/4C over 3, 5, and 12 days (Figs.~\ref{fig4}c-k; parameter settings follow Section~\ref{subsec3}, except for an imaging height of z = -51 mm and a TP smoothing intensity of 57 pixels) show that all letters become visually identifiable by day 2 for P$\mu$MA2 (125,412 TPs with sample; 426,352 TPs without sample) and by day 3 for P$\mu$MA4/4C (106,228 IPs; 53,648 TPs; 39,699 capped TPs). These observations confirm millimeter-scale resolution for medium-$Z$ material (copper, $Z=29$).

By contrast, even with the full 12-day dataset, the MSTC reconstruction cannot yield visually discernible letter patterns (Fig.~\ref{fig4}b). This degradation arises because, for moderately dense samples where average scattering angles are small, the PoCA-estimated scattering position under the single-scattering approximation can deviate substantially from the true muon trajectory in the horizontal plane, leading to significant image blurring.

\section{Conclusion}\label{sec5}

In summary, P$\mu$MA suggests an alternative formulation of scattering-sensitive particle imaging based on statistical projection responses rather than scattering-point reconstruction. By replacing conventional scattering-point estimation with transmission-track projection shifts, P$\mu$MA achieves millimeter-scale resolution---reaching a 1.196 mm knife-edge width in cosmic-ray simulations and 48 $\mu$m under idealized parallel-beam conditions---using as few as two detector planes. Experiments further demonstrate that 2-mm copper structures can be clearly resolved within two days, a performance unattainable with standard PoCA-based methods under equivalent conditions.

The ability of P$\mu$MA2 to perform high-resolution imaging with only two detectors has an profound implication. Conventional MST assumes that scattering information is encoded in reconstructed scattering locations and therefore requires independent measurements of incoming and outgoing trajectories. In contrast, the present results indicate that useful spatial information can emerge directly from the statistical redistribution of transmitted particle flow, even when scattering locations cannot be determined. Under this framework, multiple Coulomb scattering is converted into a measurable projection-shift distribution that naturally enhances edge localization at material boundaries.

Because the projection-shift principle depends only on the statistical transport behavior of penetrating particles, it is inherently extendable to other scattering-sensitive particle streams, including muon beams, protons, ions and high-energy electrons. Beyond its immediate applications in high-resolution non-destructive testing, P$\mu$MA establishes statistical projection responses as a powerful imaging observable in scattering-dominated systems where particle trajectories inside matter are not uniquely accessible.

\section*{Declarations}

\begin{itemize}
\item Funding:
This work was supported in part by the National Natural Science Foundation of China (Grant Nos. 12325504 and 12061141002) and the State Key Laboratory of Nuclear Physics and Technology, Peking University (Nos. NPT2024ZX01 and NPT2025ZX02). 
\item Patent Declaration:
A patent application related to the technologies described in this work has been published in China (Publication Number: CN 121558779 A; Publication Date: February 24, 2026).
\item Ethics approval and consent to participate:
Not applicable.
\item Consent for publication: All authors have read and approved the final version of the manuscript. We confirm that the work is original. All authors consent to its publication. 
\item Data availability:
All data included in this study are available from \url{https://zenodo.org/records/20560520}.
\item Materials availability:
Not applicable.
\item Code availability: The code involved in this study can be obtained by contacting the corresponding authors.
\item Author contribution:
Zibo Qin: Methodology, Software, Validation, Formal analysis, Investigation, Writing -- Original Draft, Visualization. 
Qite Li: Conceptualization, Methodology, Resources, Writing -- Review \& Editing, Supervision. 
Rongfeng Zhang: Software, Formal analysis, Visualization. 
Pei Yu: Software, Visualization, Writing -- Review \& Editing. 
Cheng-en Liu: Investigation. 
Liangwen Chen: Writing -- Review \& Editing. 
Feng Zhang: Writing -- Review \& Editing. 
Qiang Li: Resources, Writing -- Review \& Editing, Supervision, Project administration. 
\end{itemize}

\begin{appendices}

\section{Fundamental Formulas of Cosmic-Ray Muon Transmission and Scattering Processes and Their Application in the P$\mu$MA Methods}\label{secA1}

When cosmic-ray muons traverse matter, their interactions are dominated by two physical processes: (1) energy loss and (2) multiple Coulomb scattering (MCS). The mean energy loss rate over distance, or mass stopping power $\langle-d E / d X\rangle$, is described by the Bethe-Bloch formula \cite{Zyla2020}:
\begin{equation}
\left\langle-\frac{d E}{d X}\right\rangle=K z^{2} \frac{Z}{A} \frac{1}{\beta^{2}}\left[\frac{1}{2} \ln \frac{2 m_{e} c^{2} \beta^{2} \gamma^{2} W_{\max }}{I^{2}}-\beta^{2}-\frac{\delta(\beta \gamma)}{2}\right], \label{eqA1}
\end{equation}
where $K$ is a constant, $z$ is the charge number of the incident particle (for muon $z=1$), $Z$ and $A$ are the atomic number and mass of the target, $\beta$ is the relativistic velocity $v/c$ of the incident particle, $m_{e}$ is the electron rest mass, $\gamma$ is the Lorentz factor $\left(1-\beta^{2}\right)^{-1 / 2}$, $I$ is the mean excitation energy, $W_{\max }$ is the maximum energy transfer in a single collision, and $\delta(\beta \gamma)$ is the density effect correction to ionization energy loss. Eq.~\ref{eqA1} provides an accurate description for $0.1 \leq \beta\gamma \leq 1000$ in medium-$Z$ materials. 

Simultaneously, for monochromatic muons, MCS produces an approximately Gaussian angular distribution. The root-mean-square scattering angle is given by \cite{Bethe1953}: 
\begin{equation}
\sigma_{\theta}=\frac{13.6 \mathrm{MeV}}{\beta c p} z \sqrt{\frac{x}{X_{0}}}\left[1+0.038 \ln \left(\frac{x z^{2}}{X_{0} \beta^{2}}\right)\right], \label{eqA2}
\end{equation}
where $p$ is the momentum of the incident particle, $x$ is the target thickness, and $X_{0}$ is the radiation length, which is given by: 
\begin{equation}
X_{0}=716.4 \mathrm{~g} / \mathrm{cm}^{2} \frac{A}{Z(Z+1) \ln (287 / \sqrt{Z})}. \label{eqA3}
\end{equation}
Other parameters have the same meanings as in Eq.~\ref{eqA1}.

The horizontal deviation of the transmission point (TP) from the intersection of the true muon track in the imaging plane can be estimated using a simplified single-scattering model, in which the scattering point is defined by the intersection of the incident and outgoing tracks (assuming they are coplanar). For the P$\mu$MA2 configuration, $h_1$ and $h_2$ denote the distances from the imaging plane to the upper and lower detector layers, respectively. For P$\mu$MA4, these parameters correspond to average upstream and downstream distances, while $l_1$ and $l_2$ represent the separations within each detector group. Because the sample thickness is small compared with $h_1$ and $h_2$, the distance between the scattering point and imaging plane is neglected. 

A spherical coordinate system $(r,\theta,\phi)$ is introduced for each event, with the scattering point at the origin. The incident direction is fixed at $\theta=\pi$, and the outgoing direction at $(\pi-\theta_s,\phi_s)$, where $\theta_s$ and $\phi_s$ are the scattering and azimuthal angles. The zenith angle of the incident muon is $\alpha$, and the detector $z$-axis corresponds to $(\theta,\phi)=(\alpha,0)$; when $\alpha=0$, the two coordinate systems become parallel.

The constructed TP deviation $d$ in the imaging plane is then 
\begin{equation}
d = H\Theta(\alpha, \theta_s, \phi_s), \label{eqA4}
\end{equation}
where 
for P$\mu$MA2:
\begin{equation}
H = \frac{h_1 h_2}{h_1 + h_2}, \label{eqA5}
\end{equation}
for P$\mu$MA4:
\begin{equation}
H = \frac{4 h_1 h_2 (h_1 + h_2) + l_2 ^ 2 h_1 +l_1 ^ 2 h_2}{4 (h_1 + h_2) ^ 2 + 2 (l_1 ^ 2 + l_2 ^ 2)}, \label{eqA6}
\end{equation}
and
\begin{equation}
\Theta(\alpha, \theta_s, \phi_s)=\frac{\sin \theta_s \sqrt{1 + \tan^2 \alpha \cos^2 \phi_s}}{\cos \theta_s \cos \alpha + \sin \theta_s \sin \alpha \cos \phi_s}. \label{eqA7}
\end{equation}

For small $\theta_s$, a first-order expansion yields 
\begin{equation}
d = H \frac{\sqrt{1 + \tan^2 \alpha \cos^2 \phi_s}}{\cos \alpha} \theta_s + O({\theta_s}^2). \label{eqA8}
\end{equation}
For $\alpha\neq 0$, the deviation becomes anisotropic, though it remains proportional to the scattering angle in the small-angle regime. 

To obtain the radial distribution of $d$ for realistic cosmic-ray muons, Eq.~\ref{eqA4} must be integrated over the distributions of $(\alpha,\theta_s,\phi_s)$, with $\theta_s$ convolved with the cosmic-ray energy spectrum. Regions of higher areal density produce lower peaks and broader variances in the radial $d$ distribution, leading to characteristic undershoot and overshoot features at density boundaries. Capping the scattering angle effectively multiplies the transmissivity by the fraction of small-angle scattering events, which decreases with increasing areal density, thereby enhancing contrast. This restriction also indirectly limits the allowed range of $d$, suppressing large deviations that would otherwise originate from high-areal-density regions, thus eliminating the undershoot-overshoot behavior.

Unlike conventional scattering methods, which determine the scattering point from two independently reconstructed tracks, the P$\mu$MA methods jointly utilize upstream and downstream position measurements. This approach improves horizontal localization for the dominant small-angle population and reduces sensitivity to detector-resolution limitations. In contrast, PoCA-based reconstruction becomes increasingly unstable at small scattering angles, where minute detector uncertainties can produce large shifts in the inferred scattering point. As a result, P$\mu$MA methods provide substantially improved horizontal resolution relative to traditional scattering-based techniques. 

\section{Details of the P$\mu$MA framework}\label{secA2}

The P$\mu$MA framework includes many specific configurations. The simplest configuration, P$\mu$MA2, employs only two detectors---one upstream and one downstream of the sample. Its main procedures are described below. 

\begin{enumerate}
    \item \textbf{Coincidence Logic Settings:} 
    Full coincidence is required for both the upstream and downstream detectors to reliably record valid muon events. 

    \item \textbf{Transmission Track Construction:} 
    The straight line connecting the hit positions recorded by the coincident upper and lower detectors defines the transmission track. 

    \item \textbf{Projection Shift Construction:} 
    The intersection of each transmission track with an x-y imaging plane, placed near the center of the sample along the z-axis, is defined as the Transmission Point (TP), or the P$\mu$MA point. (Refer to Section~\ref{subsec1} for the coordinate system.) For each muon, projection of the constructed transmission track onto the imaging plane converts the small material-induced angular deflection into an in-plane displacement, causing the resulting TP to appear shifted relative to its true transmission position. The imaging plane is divided into a grid, and the TP count in each cell forms the TP matrix. For P$\mu$MA2, two TP matrices are required: one with the sample and one without it.
    
    \item \textbf{Pre-smoothing:} 
    Spatial filtering is applied to both TP matrices to suppress statistical fluctuations. Mean filtering is used in this work, with a stronger kernel applied to the matrix without sample: 
    \begin{equation}
        \begin{aligned}
            \textbf{TP}_{\text{smoothed, without sample}} &= \mathcal{F}_{\text{strong}}(\textbf{TP}_{\text{raw, without sample}}); \\
            \textbf{TP}_{\text{smoothed, with sample}} &= \mathcal{F}_{\text{weak}}(\textbf{TP}_{\text{raw, with sample}}).
        \end{aligned} \label{eq1}
    \end{equation}
    The terms "strong" and "weak" refer to the spatial extent of the applied filter kernel.
    
    \item \textbf{Transmission Ratio Imaging:} 
    The Transmission Ratio (TR) matrix is then computed as
    \begin{equation}
        \mathrm{\textbf{TR}}_{ij} = \frac{[\textbf{TP}_{\text{smoothed, with sample}}]_{ij}}{[\textbf{TP}_{\text{smoothed, without sample}}]_{ij}},\label{eq2}
    \end{equation}
    where $ij$ indexes the grid cell. To control statistical errors, elements in the denominator must exceed a predefined threshold; otherwise, the corresponding TR elements are marked as invalid. When the sample placement region is known, a suitable sample-free area can be selected such that its TP counts are compared with those at the same locations in the empty-setup data, ensuring that the TR in the sample-free region is normalized to unity.
\end{enumerate}

The transmission ratio---defined here as the ratio of TPs to IPs---is not identical to the physical transmissivity. The discussion in this section pertains specifically to this TP/IP-based definition. Experimentally measured transmission ratios (even with no cut) are systematically lower than the physical transmissivity due to two dominant effects: 

\begin{enumerate}
\item \textbf{Geometric acceptance:}
Some outgoing tracks fall outside the sensitive region of the downstream detectors, even when their corresponding incident tracks intersect it.
\item \textbf{Detector efficiency:}
Some outgoing tracks are not registered despite passing through the sensitive area, because the detector efficiency is below 100\%.
\end{enumerate}

In simulated data, only geometric acceptance applies, leading to higher transmission ratios than those measured experimentally. This difference does not affect the comparison of materials’ muon-stopping capabilities under identical conditions. However, the limited efficiency of the experimental detectors increases the exposure time required to obtain images of sufficient quality.

When four detectors are available (P$\mu$MA4), arranged as two upstream and two downstream, the upstream group provides incident-track information that serves as a background reference. Several differences arise compared with P$\mu$MA2. The upstream detectors operate with an independent coincidence trigger to record particles that do not traverse the sample region. The incident track is defined as the straight line extrapolated from the upstream hit positions, and its intersection with the imaging plane is recorded as the Incident Point (IP). These IPs are accumulated into an IP matrix analogous to the TP matrix. If the incident track does not intersect the active area of each downstream detector, the corresponding IP is discarded. The transmission track is obtained by fitting a 3D straight line to coincident hits across all detectors using a least-squares method that treats the z coordinate as the independent variable. Subsequent processing mirrors P$\mu$MA2, with the IP matrix replacing the TP matrix without sample. Alternatively, as in P$\mu$MA2, a TP matrix acquired without the sample can also be used as a background reference.

A variant, P$\mu$MA4C, retains the same transmission-ratio construction but additionally computes a scattering angle for each TP event. A maximum allowed scattering angle $\theta_{\max}$ is imposed, with $\theta_{\max}$ chosen to be on the same order as the standard deviation of the scattering angle in the no-sample configuration. Events exceeding $\theta_{\max}$ are excluded from the TP count. This selection preferentially retains tracks close to straight-line propagation, mitigating image blurring caused by large-angle deviations and improving contrast between regions of differing areal density. By capping the scattering angle, the effective transmissivity is multiplied by the fraction of small-angle scattering events relative to the total, which decreases with increasing areal density and thereby enhances contrast. 

\section{Image Quality Assessment Metrics and Calculation Methods Employed in This Study}\label{secA3}

\subsection{The knife-edge width}\label{subsecA1}

The 20-80\% method for knife-edge width measurement, adapted from rise-time characterization in electronic signal processing \cite{Paulter2004}, is defined as follows: 

\begin{itemize}

\item The total edge variation $\Delta$ is the peak-to-peak value of the edge spread function (ESF) near the edge:
\begin{equation}
\Delta = |\text{ESF}_{\text{max}} - \text{ESF}_{\text{min}}|. \label{eqC1}
\end{equation}
In this study, the region within 20 mm from the edge is used to evaluate $\Delta$.

\item The knife-edge width $w$ is the distance between the positions $x_{20}$ and $x_{80}$ that satisfy
\begin{equation}
\begin{aligned}
\text{ESF}(x_{20}) &= \text{ESF}_{\text{min}} + 0.2\Delta, \\
\text{ESF}(x_{80}) &= \text{ESF}_{\text{min}} + 0.8\Delta,
\end{aligned} \label{eqC2}
\end{equation}
yielding
\begin{equation}
w = |x_{80} - x_{20}|. \label{eqC3}
\end{equation}

\end{itemize}

The knife-edge width provides a practical measure of image acutance: a smaller width indicates a sharper edge transition. A reduced knife-edge width also implies a narrower and steeper line spread function (LSF), the derivative of the ESF, which in turn leads to improved spatial resolution as determined from the modulation transfer function (MTF) (see Section~\ref{subsecA2}). 

\subsection{The spatial resolution}\label{subsecA2} 

Spatial resolution is quantified using the modulation transfer function (MTF) method \cite{Maitre2017}. A precision knife-edge target is placed at a small angle ($2^{\circ}$ in this work) relative to the pixel matrix to enable sub-pixel sampling of the edge spread function (ESF). The ESF is constructed from intensity profiles perpendicular to the edge, and numerical differentiation yields the line spread function (LSF) via $\text{LSF}(x) = d(\text{ESF})/dx$.

The MTF curve is computed as the normalized magnitude of the Fourier transform of the LSF:
\begin{equation}
\text{MTF}(f) = \frac{|F\{\text{LSF}(x)\}|} {|F\{\text{LSF}(0)\}|},  \label{eqC4}
\end{equation}
where $f$ denotes spatial frequency in line pairs per millimeter (lp/mm). The spatial resolution is then defined as the reciprocal of the spatial frequency at which the MTF falls to a specified fraction of its zero-frequency value, typically $f_{\text{MTF50}}$ for $\text{MTF}=0.5$ or $f_{\text{MTF10}}$ for $\text{MTF}=0.1$. This method provides an objective and system-independent measure of imaging performance suitable for comparison across different modalities.

\subsection{The peak signal-to-noise ratio}\label{subsecA3}

The peak signal-to-noise ratio (PSNR) is widely used to quantify image quality by comparing the maximum signal level to the noise power, expressed in decibels (dB) \cite{Maitre2017}. For an $m \times n$ image, the PSNR between a reference image $I$ and a noisy approximation $K$ is defined as

\begin{equation}
\text{PSNR} = 10 \cdot \log{10}\left(\frac{\text{MAX}^2}{\text{MSE}}\right) \text{[dB]}, \label{eqC5}
\end{equation}

where:

\begin{itemize}

\item MAX is the maximum possible pixel value (e.g., 255 for 8-bit, 65535 for 16-bit images);

\item MSE is the mean squared error,

\begin{equation}
\text{MSE} = \frac{1}{mn}\sum_{i=0}^{m-1}\sum_{j=0}^{n-1}\left[I(i,j) - K(i,j)\right]^2. \label{eqC6}
\end{equation}

\end{itemize}

In this study, the PSNR calculation is reduced to one dimension due to the presence of a long, straight edge. MAX is defined as the ESF range. In the absence of a noise-free reference, MSE is estimated from regions located 20-40 mm away from the edge, where the ESF exhibits almost linear behavior. A linear fit is applied to these regions, and the MSE is approximated by the root-mean-square of the residuals.

One-dimensional PSNR values tend to be higher than those derived from two-dimensional images. Nevertheless, when identical ESF sampling procedures and datasets are used, the resulting PSNR values remain directly comparable across different imaging methods.

\section{The Areal Density Response of P$\mu$MA Methods}\label{secA4}

As transmission-based imaging techniques, the P$\mu$MA methods provide not only horizontal spatial resolution but also sensitivity to variations in areal density. Regions with different areal densities appear with distinct gray levels in the reconstructed images. This section examines the areal-density response of the P$\mu$MA4 and P$\mu$MA4C methods using simulated data.

\begin{figure}[H]
\centering
\includegraphics[width=1.0\textwidth]{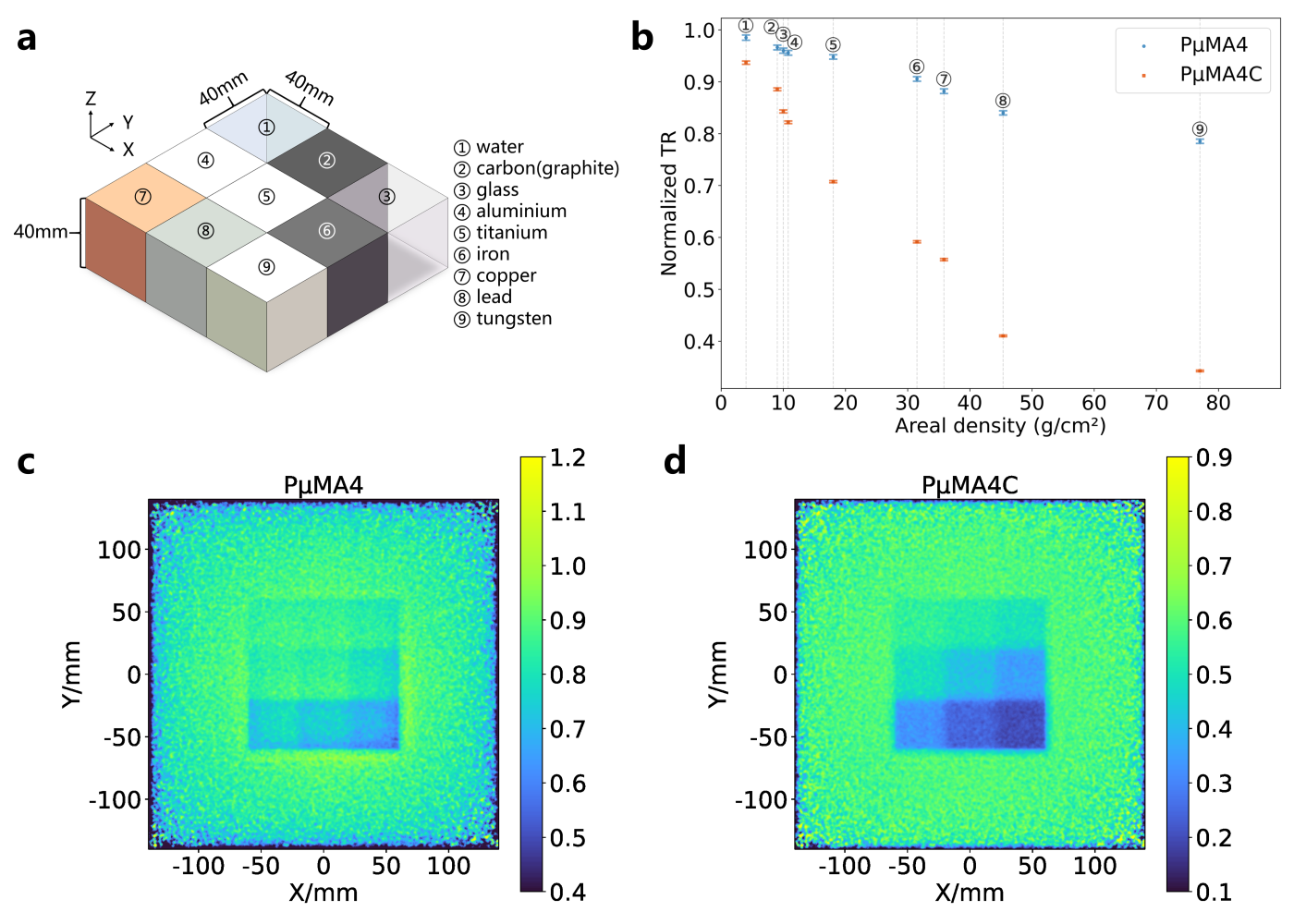}
\caption{\textbf{a}, Schematic of the sample composed of nine materials. Each block measures 40 mm × 40 mm × 40 mm. The geometric center of the central block (No. 5) is located at (x, y, z) = (0, 0, -58 mm), and the coordinate system is indicated in the diagram. \textbf{b}, Transmission ratio (TR) values of the nine materials obtained with the P$\mu$MA4 and P$\mu$MA4C methods, normalized to the air TR (set to 1). \textbf{c}, Simulated reconstruction using the P$\mu$MA4 method. \textbf{d}, Simulated reconstruction using the P$\mu$MA4C method.}\label{fig5}
\end{figure}

The simulation setup follows Section~\ref{subsec2}. A relatively large dataset (9,667,336 IPs; 8,042,994 TPs; 4,839,405 post-capping TPs) was generated to suppress statistical noise. The sample configuration is shown in Fig.~\ref{fig5}a and its caption, and the imaging parameterization matches that in Section~\ref{subsec3}. Under these low-noise conditions, a 20 mm × 20 mm region centered on each material block is used to compute a preliminary transmission ratio (TR). Each TR value is then normalized by the TP-IP ratio obtained from an air-only simulation of comparable statistics within the same region. The resulting normalized TRs are shown in Fig.~\ref{fig5}b, while reconstructed images by P$\mu$MA4 and P$\mu$MA4C appear in Figs.~\ref{fig5}c and \ref{fig5}d.

The normalized TRs from P$\mu$MA4C show enhanced sensitivity to changes in areal density, demonstrating stronger discrimination capability. In contrast, the normalized TRs obtained with P$\mu$MA4 exhibit limited separation among several medium- and low-density materials (graphite, glass, aluminum, titanium). Nevertheless, the wrinkle-like contrast patterns visible in Fig.~\ref{fig5}c still indicate abrupt changes in areal density.

As expected, distinguishing two regions becomes more difficult as their areal-density difference decreases, since the signal reflecting their contrast (e.g., the TR difference) must significantly exceed the noise level to be visually or algorithmically detectable. For large contrasts---such as several centimeters of air versus lead---clear differentiation can emerge within only a few hours of cosmic-ray exposure. However, for extremely small areal-density variations, the limited flux of cosmic rays renders reliable discrimination impractical. This regime likely represents a natural domain for high-collimation muon-beam transmission imaging, where substantially improved statistics can be achieved.




\end{appendices}



\begin{thebibliography}{99}

\bibitem{Zyla2020}
Zyla, P.~A. et al. (Particle Data Group). Review of Particle Physics. \emph{Prog. Theor. Exp. Phys.} \textbf{2020}, 083C01 (2020).

\bibitem{Borozdin2003}
Borozdin, K.~N. et al. Radiographic imaging with cosmic-ray muons. \emph{Nature} \textbf{422}, 277 (2003).

\bibitem{Bonomi2020}
Bonomi, G., Checchia, P., D'Errico, M., Pagano, D. \& Saracino, G. Applications of cosmic-ray muons. \emph{Prog. Part. Nucl. Phys.} \textbf{112}, 103768 (2020).

\bibitem{Alvarez1970} 
Alvarez, L.~W. et al. Search for Hidden Chambers in the Pyramids. \emph{Science} \textbf{167}, 832--839 (1970).

\bibitem{Morishima2017}
Morishima, K. et al. Discovery of a big void in Khufu's Pyramid by observation of cosmic-ray muons. \emph{Nature} \textbf{552}, 386--390 (2017).

\bibitem{Nagamine1995}
Nagamine, K., Iwasaki, M., Shimomura, K. \& Ishida, K. Method of probing inner-structure of geophysical substance with the horizontal cosmic-ray muons and possible application to volcanic eruption prediction. \emph{Nucl. Instrum. Methods Phys. Res., Sect. A} \textbf{356}, 585--595 (1995).

\bibitem{D'Errico2020}
D'Errico, M. et al. Muon radiography applied to volcanoes imaging: the MURAVES experiment at Mt. Vesuvius. \emph{J. Instrum.} \textbf{15}, C03014 (2020).

\bibitem{Qi2026}
Qi, S. et al. Three-dimensional density and air-rock interface reconstruction with muography: Application to the TianQin tunnel. Preprint at \url{https://arxiv.org/abs/2606.03397} (2026).

\bibitem{Valencia2025}
Valencia, J.~J. et al. Simulations of muon imaging with the LANL GMT detector for spent nuclear fuel cask content verification. \emph{J. Appl. Phys.} \textbf{138}, 184502 (2025).

\bibitem{Wen2023}
Wen, Q.~G. Research on rapid imaging with cosmic ray muon scattering tomography. \emph{Sci. Rep.} \textbf{13}, 19718 (2023).

\bibitem{Yu2024}
Yu, P. et al. A new efficient imaging reconstruction method for muon scattering tomography. \emph{Nucl. Instrum. Methods Phys. Res., Sect. A} \textbf{1069}, 169932 (2024).

\bibitem{Wu2026}
Wu, Y.~N. et al. Reinforcement learning for muon scattering tomography enhancement. \emph{Nucl. Sci. Tech.} \textbf{37}, 81 (2026).

\bibitem{Xu2025} 
Xu, Y. et al. Feasibility study of the GeV-energy muon source based on the High Intensity Heavy-Ion Accelerator Facility. \emph{Phys. Rev. Accel. Beams} \textbf{28}, 053401 (2025).

\bibitem{LiuF2025}
Liu, F. et al. Simulation studies of a high-repetition-rate electron-driven surface muon beamline at SHINE. \emph{Phys. Rev. Accel. Beams} \textbf{28}, 083401 (2025).

\bibitem{Cook2017}
Cook, S. et al. Delivering the world's most intense muon beam. \emph{Phys. Rev. Accel. Beams} \textbf{20}, 030101 (2017).

\bibitem{Zhang2025}
Zhang, F. et al. Proof-of-principle demonstration of muon production with an ultrashort high-intensity laser. \emph{Nat. Phys.} \textbf{21}, 1050--1056 (2025).

\bibitem{YuX2024}
Yu, X. et al. Proposed Peking University muon experiment for muon tomography and dark matter search. \emph{Phys. Rev. D} \textbf{110}, 016017 (2024).

\bibitem{Liu2026}
Liu, C. et al. Probing cosmic ray composition and muonphilic dark matter via muon tomography. \emph{Phys. Rev. Lett.} \textbf{136}, 151001 (2026).

\bibitem{Liu2026-2}
Liu, C. et al. Probing and knocking with muons and new physics exploration (in Chinese). \emph{Chin. Sci. Bull.} \textbf{71}, 894--903 (2026).

\bibitem{Li2012}
Li, Q. et al. Study of spatial resolution properties of a glass RPC. \emph{Nucl. Instrum. Methods Phys. Res., Sect. A} \textbf{663}, 22--25 (2012).

\bibitem{Li2013}
Li, Q.~T. et al. A sub-millimeter spatial resolution achieved by a large sized glass RPC. \emph{Chin. Phys. C} \textbf{37}, 016002 (2013).

\bibitem{Chen2014}
Chen, S. et al. Simulation of a small muon tomography station system based on RPCs. \emph{J. Instrum.} \textbf{9}, C10022 (2014).

\bibitem{Agostinelli2003}
Agostinelli, S. et al. Geant4---a simulation toolkit. \emph{Nucl. Instrum. Methods Phys. Res., Sect. A} \textbf{506}, 250--303 (2003).

\bibitem{Allison2006}
Allison, J. et al. Geant4 developments and applications. \emph{IEEE Trans. Nucl. Sci.} \textbf{53}, 270--278 (2006).

\bibitem{Allison2016}
Allison, J. et al. Recent developments in Geant4. \emph{Nucl. Instrum. Methods Phys. Res., Sect. A} \textbf{835}, 186--225 (2016).

\bibitem{PKMUON2024} 
PKMuon Collaboration. PKMUON\_2024: newrpc.yaml configuration file. 
\url{https://github.com/PKMuon/PKMUON_2024/blob/text/config/newrpc.yaml} (2025).

\bibitem{Hagmann2007}
Hagmann, C., Lange, D. \& Wright, D. Cosmic-ray shower generator (CRY) for Monte Carlo transport codes. \emph{2007 IEEE Nuclear Science Symposium Conference Record} 1143--1146 (IEEE, 2007). 

\bibitem{Bethe1953}
Bethe, H.~A. Molière's Theory of Multiple Scattering. \emph{Phys. Rev.} \textbf{89}, 1256--1266 (1953).

\bibitem{Paulter2004}
Paulter, N.~G., Larson, D.~R. \& Blair, J.~J. The IEEE Standard on Transitions, Pulses, and Related Waveforms, Std-181-2003. \emph{IEEE Trans. Instrum. Meas.} \textbf{53}, 1209--1217 (2004).

\bibitem{Maitre2017}
Maître, H. Image Quality. In \emph{From Photon to Pixel} (ed. Maître, H.) 205--255 (Wiley, 2017).

\end{thebibliography}

\end{document}